\documentclass[12pt]{article}
\usepackage{graphicx}
\newcommand{\kms}{$\rm{km\,s^{-1}}$} 
\newcommand{\mic}{$\mu$m}

\newcommand{\irnulvier}{IRAS\,04296+3429}
\newcommand{\irnulvijf}{IRAS\,05341+0852}
\newcommand{\irnulzeven}{IRAS\,07134+1005}
\newcommand{\ireennegen}{IRAS\,19500$-$1709}
\newcommand{\irtweetwee}{IRAS\,22223+4327}
\newcommand{\irtweedrie}{IRAS\,23304+6147}

\title{\bf Probing S-Process Yields of Field Carbon Stars by the Analysis
of 21\mic\ Post-AGB Stars\footnote{based on observations collected at 
the European Southern Observatory in Chile (61.E-0426),
and at Roque de los Muchachos at La Palma, Spain}}
\author{M.~Reyniers $^1$\thanks{Scientific 
researcher of the Fund for Scientific Research, Flanders} \\ 
and H.~Van~Winckel $^1$\thanks{Postdoctoral fellow of the Fund
for Scientific Research, Flanders}\\
\vspace{1cm}\\
\normalsize $^1$Instituut voor Sterrenkunde, K.U.Leuven, Celestijnenlaan 200B, 
B-3001 Leuven, Belgium}
\date{\mbox{}}
\begin{document}
\maketitle
\pagestyle{empty}
%
%
\def\bull{\vrule height .9ex width .8ex depth -.1ex}
\makeatletter
\def\ps@plain{\let\@mkboth\gobbletwo
\def\@oddhead{}\def\@oddfoot{\hfil\tiny\bull\quad
``The Galactic Halo: from Globular Clusters to Field Stars'';
35$^{\mbox{\rm rd}}$ Li\`ege\ Int.\ Astroph.\ Coll., 1999\quad\bull}%
\def\@evenhead{}\let\@evenfoot\@oddfoot}
\makeatother
%
%
\def\beginrefer{\section*{References}%
\begin{quotation}\mbox{}\par}
\def\refer#1\par{{\setlength{\parindent}{-\leftmargin}\indent#1\par}}
\def\endrefer{\end{quotation}}
%
%
{\noindent\small{\bf Abstract:} 
   The observational data guiding the theoretical chemical evolutionary
models of AGB stars come mainly from the analysis of intrinsic and extrinsic
s-process enriched objects. The first are the stars of the M-MS-S-SC-C star 
sequence which is thought to reflect, at least partly, 
the evolution on the AGB of a single star towards an increasing C/O ratio. 
The second are all binaries with excess abundances acquired by 
mass-transfer from the companion which is now a white dwarf.

   In this contribution we present a homogeneous abundance analysis of 21\mic\
objects, which are carbon-rich probably single post-AGB stars.
With their F to G spectral
types a wide variety of chemical species can be studied quantitatively 
by using photospheric atomic lines. These objects are among the
most s-process enriched objects known so far. We focus on the s-process
distribution and the related neutron irradiation and compare the results 
with the enrichments observed in both intrinsic and extrinsic objects.}
%
%

 \begin{figure*}
 \begin{center}
 \includegraphics[scale=.45,angle=-90]{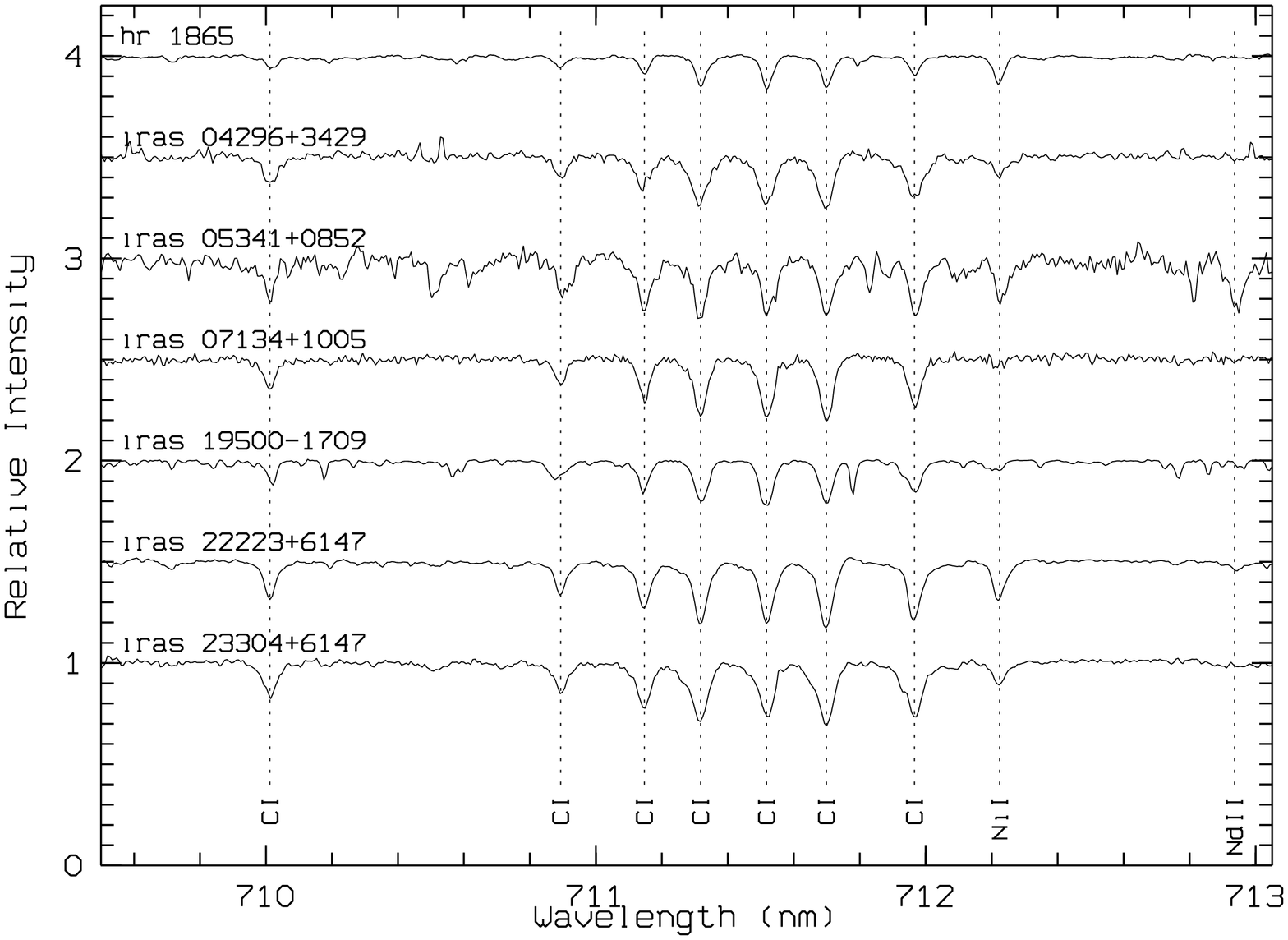}
 \end{center}
 \caption{\label{fig:carbNM}
 Sample spectra of the programme stars and HR\,1865 around the red carbon
 multiplet. The stars are velocity corrected. On top, the spectrum of the
 reference star HR\,1865 is plotted. This massive supergiant (FI0) has
 similar atmospheric parameters as the programme stars (Decin et al. 1998)
 but obviously no enrichment of helium burning products.}
 \begin{center}
 \includegraphics[scale=.45,angle=-90]{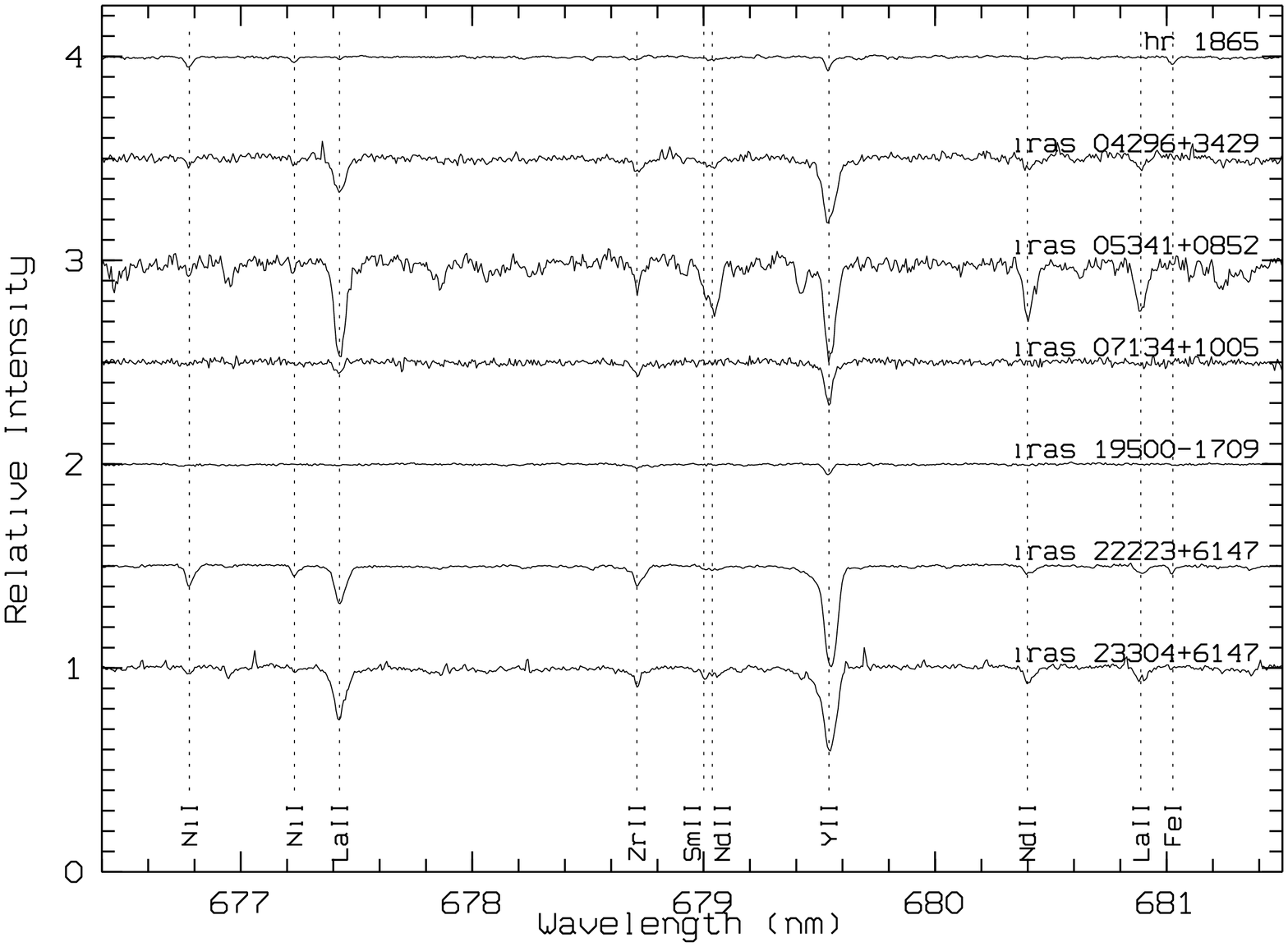}
 \end{center}
 \caption{\label{fig:sproNM}
 The region around $\lambda$679\,nm for the six programme stars and HR\,1865.
 In this region, both lines of light (ls) and heavy (hs) s-process elements
 are present. Comparison with the normal supergiant
 HR\,1865 proves that our stars are clearly s-process enhanced. The ratio
 of the line strength of the LaII-line at $\lambda$667.427\,nm and the
 YII-line at $\lambda$679.541\,nm for each star gives a very qualitative
 idea of the neutron exposure. From this spectral comparison we can
 expect a high neutron exposure for \irnulvijf.}

 \end{figure*}

\section{Introduction}
A subgroup of field post-AGB stars shows in their mid-infrared spectra a
21\mic\ emission feature. This feature is not found in AGB stars, nor in PNe,
suggesting that the excitation and/or formation of its carrier is limited to 
the dust enveloppes of the transition objects. Although the nature of the 
21\mic\ feature is not known, a good correlation exists between the presence of
circumstellar carbon molecules like C$_2$ and C$_3$ in the optical spectra and
the presence of the IR feature.

Several studies were dedicated to reveal the chemical photospheric composition
of the 21\mic\ stars, confirming their post-3rd dredge-up character (large
C/O ratios and s-process enhancements). Together with their carbon-rich dust
enveloppes, spectral types and high luminosities, the 21\mic\ stars are good
candidate {\em post-carbon} stars. Since different authors use different line 
lists and atomic data, results are difficult to compare quantitatively.
Therefore, a {\em homogeneous} (same method and line list) abundance analysis 
is presented in this contribution.

\section{Observations and Analysis}

Spectra were obtained using the Utrecht Echelle Spectrograph (UES) mounted
on the 4.2m William Hershell Telescope (WHT) during several runs between
1992 and 1995.
For one object (IRAS\,07134), the NTT was used in combination of the EMMI
spectrograph. For most spectra, we obtained a resolution of $\sim$50,000
and a S/N$>$100 (see Fig.~\ref{fig:carbNM}\ and \ref{fig:sproNM} for sample
spectra).
Model atmospheric parameters (Table~\ref{tab:modparam}) were derived
by the commonly used spectroscopic method: the effective temperature
T$_{\rm eff}$ from the excitation balance of the FeI-lines, the surface gravity
$\log(g)$ from ionisation balance between FeI and FeII and the microturbulence
$\xi_t$ by forcing the Fe abundance to be independent of the reduced equivalent
width W$_{\lambda}/\lambda$
of the FeI lines. Abundances were obtained with the Kurucz model
atmospheres. Special attention was paid to accurate $\log(gf)$ values, using
the most recent compilations. Only weak lines with equivalent widths smaller
than 150\,m\AA\ were used in the analysis.

 \begin{table}
 \caption{Model parameters and metallicity of the six programme 
 stars}\label{tab:modparam}
 \begin{center}
 \begin{tabular}{|l|ccc|c|}
 \hline
        & T$_{\rm eff}$ & $\log g$ & $\xi_t$  &  [Fe/H]\\
        &     (K)   &          &    (\kms)&  \\
 \hline
 \irnulvier  & 7000          & 1.0      & 4.0 & $-$0.6  \\
 \irnulvijf  & 6500          & 1.0      & 3.5 & $-$0.8  \\
 \irnulzeven & 7250          & 0.5      & 5.0 & $-$1.0  \\
 \ireennegen & 8000          & 1.0      & 6.0 & $-$0.6  \\
 \irtweetwee & 6500          & 1.0      & 5.5 & $-$0.3  \\
 \irtweedrie & 6750          & 0.5      & 3.0 & $-$0.8  \\
 \hline
 \end{tabular}
 \end{center}
 \end{table}

\section{Results}
\begin{figure*}
\resizebox{\hsize}{!}{\includegraphics{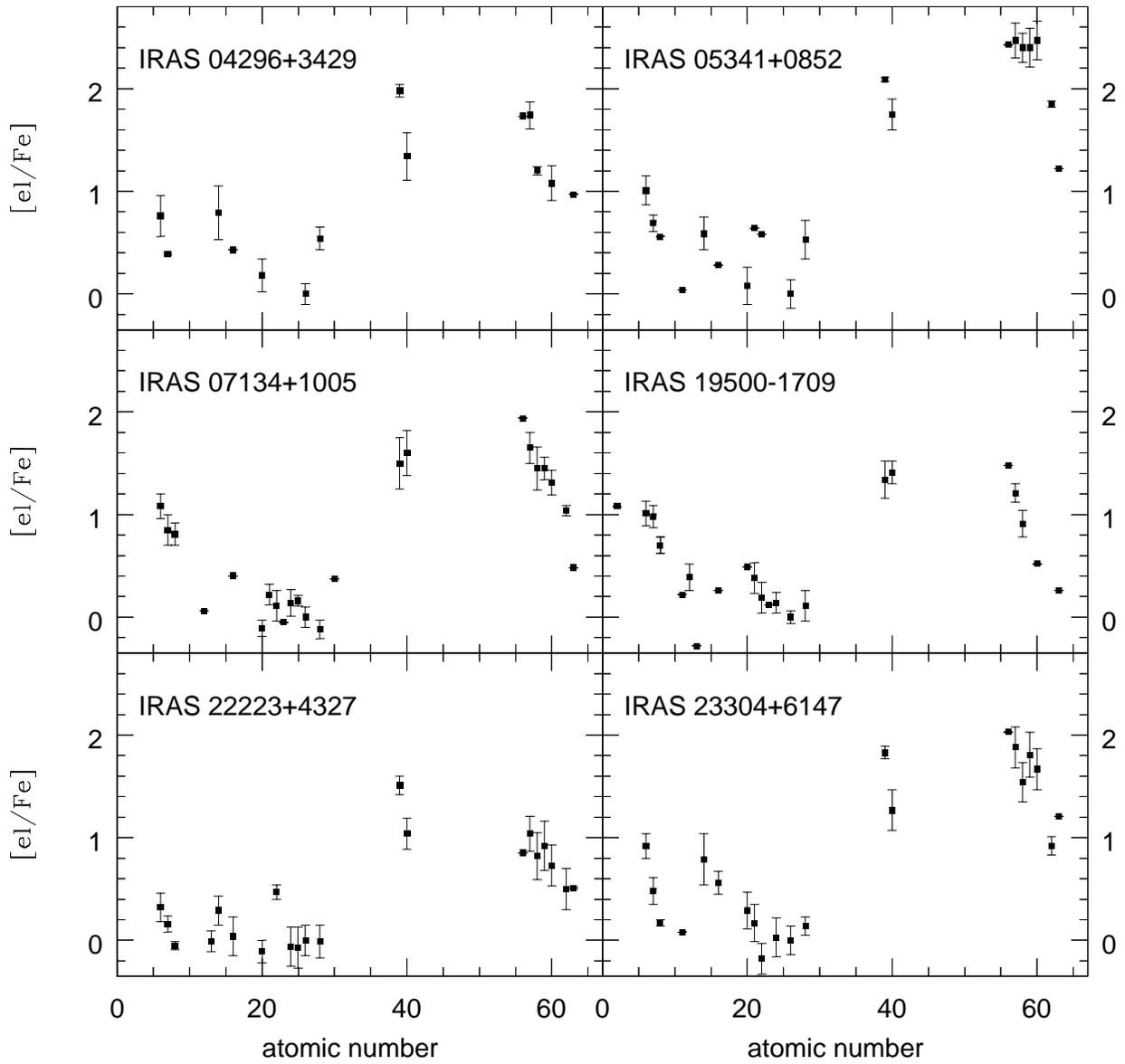}}
\caption{\label{fig:zesgraf}
The abundances of our six programme stars relative to iron [el/Fe].}
\end{figure*}

The results of our analysis are presented in Fig.~\ref{fig:zesgraf}.
The {\bf metallicity} relative to the solar value ranges from $-$0.3 to $-1.0$
(Table~\ref{tab:modparam}), indicating that our sample indeed consists of an 
old and hence low-mass population.
From Fig.~\ref{fig:zesgraf} it is clear that all stars display a huge 
photospheric {\bf carbon} enhancement. This is illustrated
in Fig.~\ref{fig:carbNM}. {\bf Oxygen} is also enhanced, but, 
especially for the cooler stars, an accurate oxygen abundance is 
hampered by the small number 
of lines. Accurate C/O ratios are therefore hard to obtain, but are found in 
between 1.0 and 2.9.
The slight overabundances of the {\bf $\alpha$-elements} are considered to be 
normal for this metallicity range.
The most convincing argument for the post-3rd dredge-up character of the 
studied stars is the huge enhancement in {\bf s-process elements}. This is
illustrated in Fig.~\ref{fig:sproNM}. The 21\mic\ stars turn out to be among 
the most s-process enriched objects known so far.

\section{Discussion}
\subsection{Neutron Exposure}
We defined for each star the [hs/ls] index as the ratio of heavy
(Ba, La, Nd, Sm) to light (Y, Zr) s-process elements. The abundances of 
unobserved species were estimated using the tables of Malaney (1987).
The [hs/ls] index is a good indicator for the {\em mean neutron exposure
$\tau_0$}, whereas [s/Fe] is a measure for the {\em total s-process enrichment}.

\begin{figure*}
 \includegraphics[scale=.5]{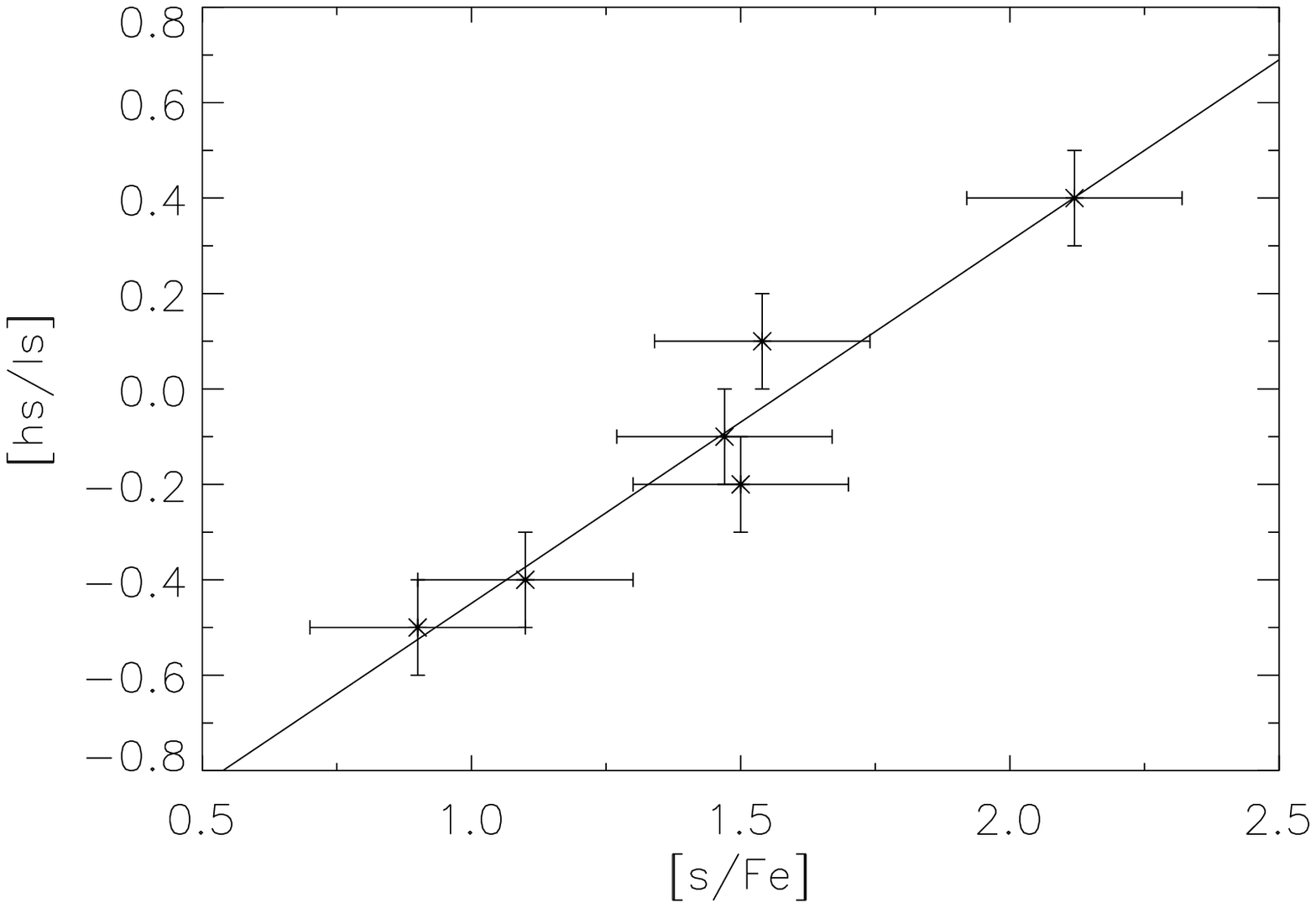}
 \includegraphics[scale=.5]{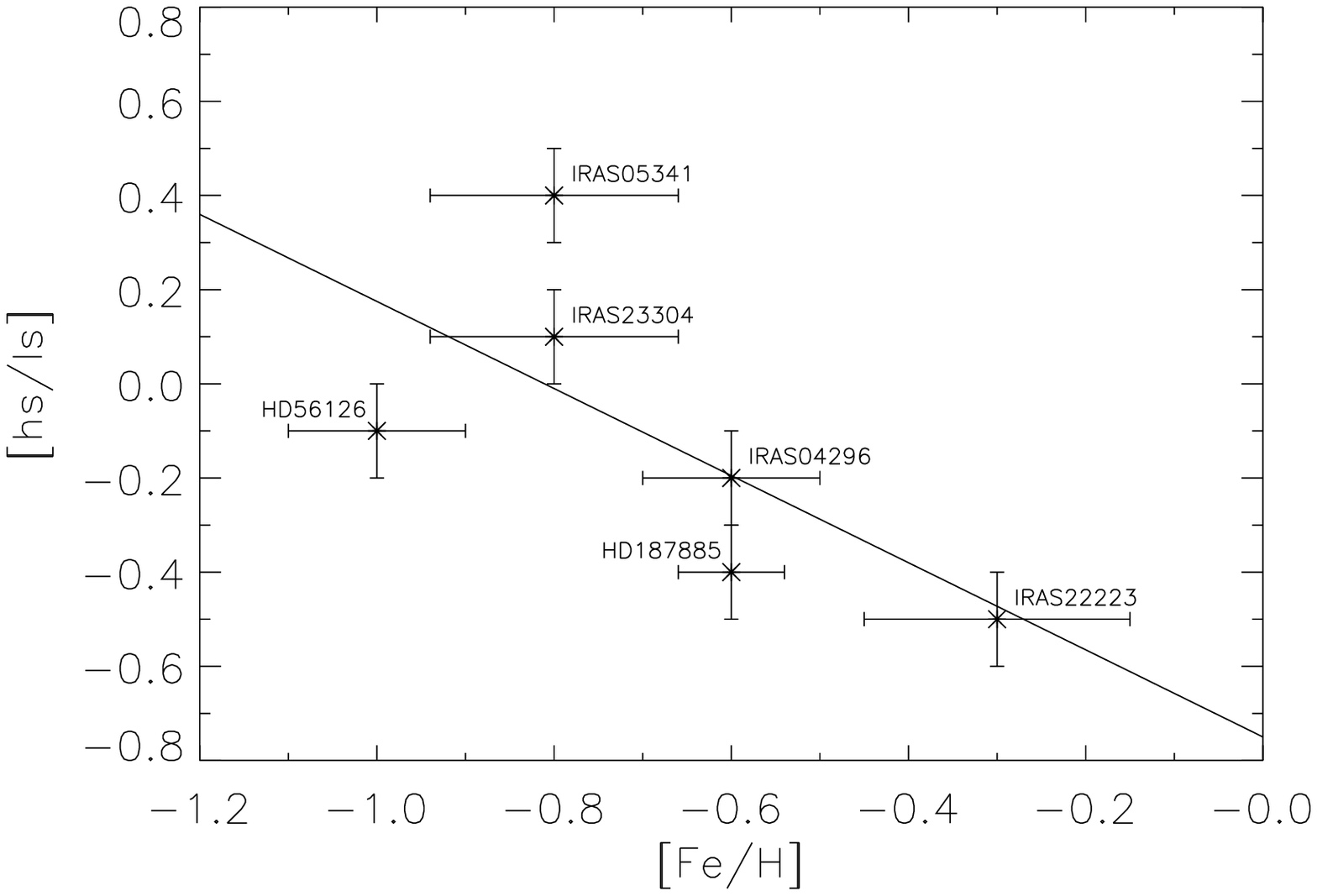}
 \caption{\label{fig:twopan}
 The [hs/ls] index as a function of the total s-process enrichment
 [s/Fe] {\em (left panel)} and as a function of metallicity [Fe/H]
 {\em (right panel).}}
 \end{figure*}

In Fig. \ref{fig:twopan} (left panel) a strong correlation is found 
between the [hs/ls] index and the total enrichment of 
the s-process elements as parameterised by the [s/Fe]-index, 
in the sense that more enriched objects also display a higher integrated 
neutron irradiation. Since in carbon stars the asymptotic values of the 
s-process distribution is probably reached (Busso et al. 1995), this means 
that the dredge-up efficiency is strongly linked with the neutron production
in the intershell!  

On the right panel of Fig. \ref{fig:twopan} the [hs/ls] index is shown as a 
function of metallicity. The expected trend (increasing [hs/ls] with decreasing
metallicity because of the larger amount of neutrons per seed nucleus) shows a
large intrinsic scatter and suggests that other fundamental parameters strongly
determine the internal nucleosynthesis and dredge-up phenomena. The 21\mic\
stars are the only intrinsically enriched objects showing a large spread
in metallicity.

\subsection{Comparison with Other Enriched Stars}
\begin{figure*}
\begin{minipage}{8cm}
\resizebox{\hsize}{!}{\includegraphics{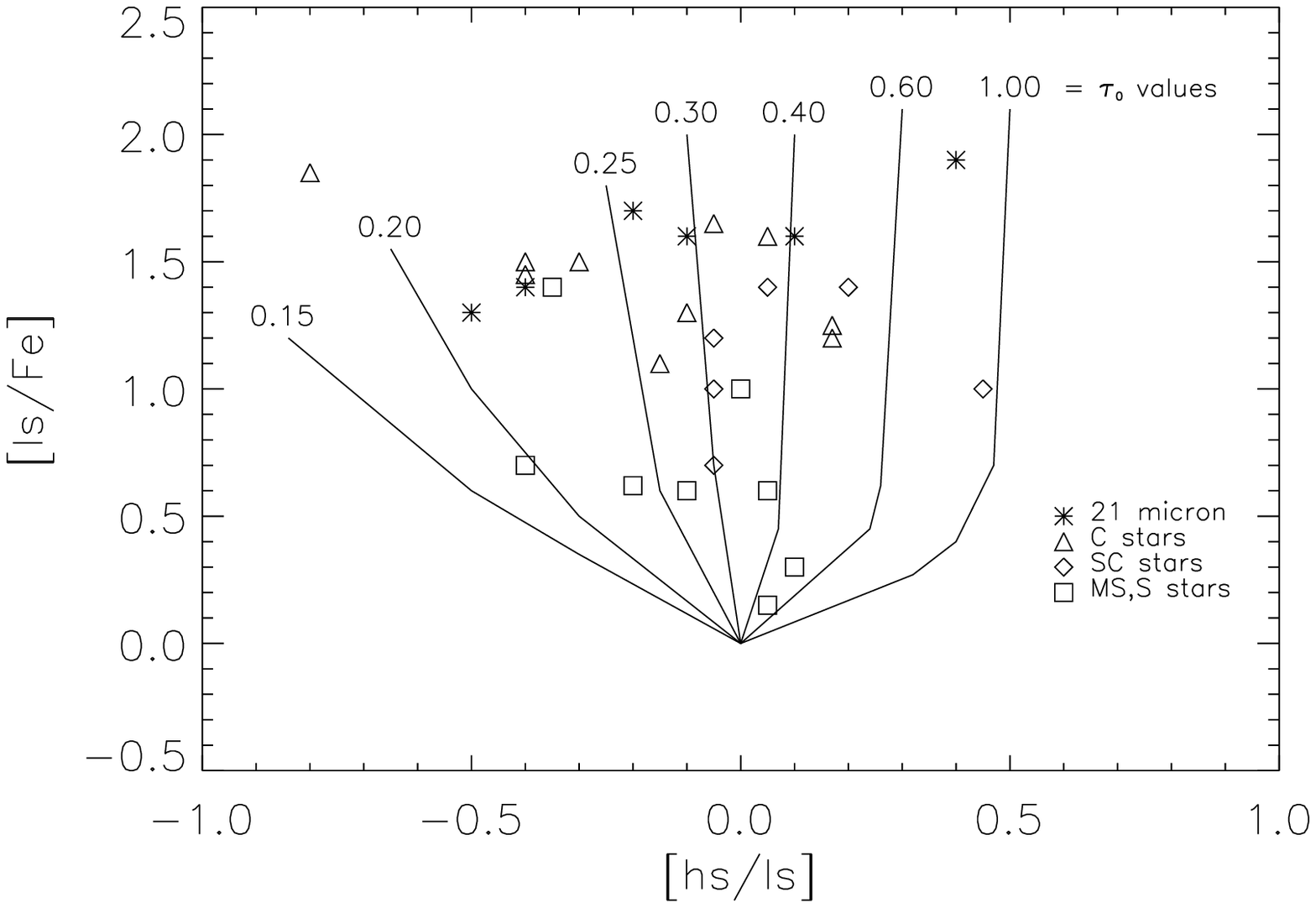}}
\caption{\label{fig:handmalaney}
The enrichment of light s-process elements [ls/Fe] as a function of the 
neutron exposure [hs/ls]. The full lines are the theoretical predictions
for different neutron exposures parameterised by $\tau_0$ from Busso et al.
(1995). Data points: see references.}
\end{minipage}
\hfill
\begin{minipage}{8cm}
\resizebox{\hsize}{!}{\includegraphics{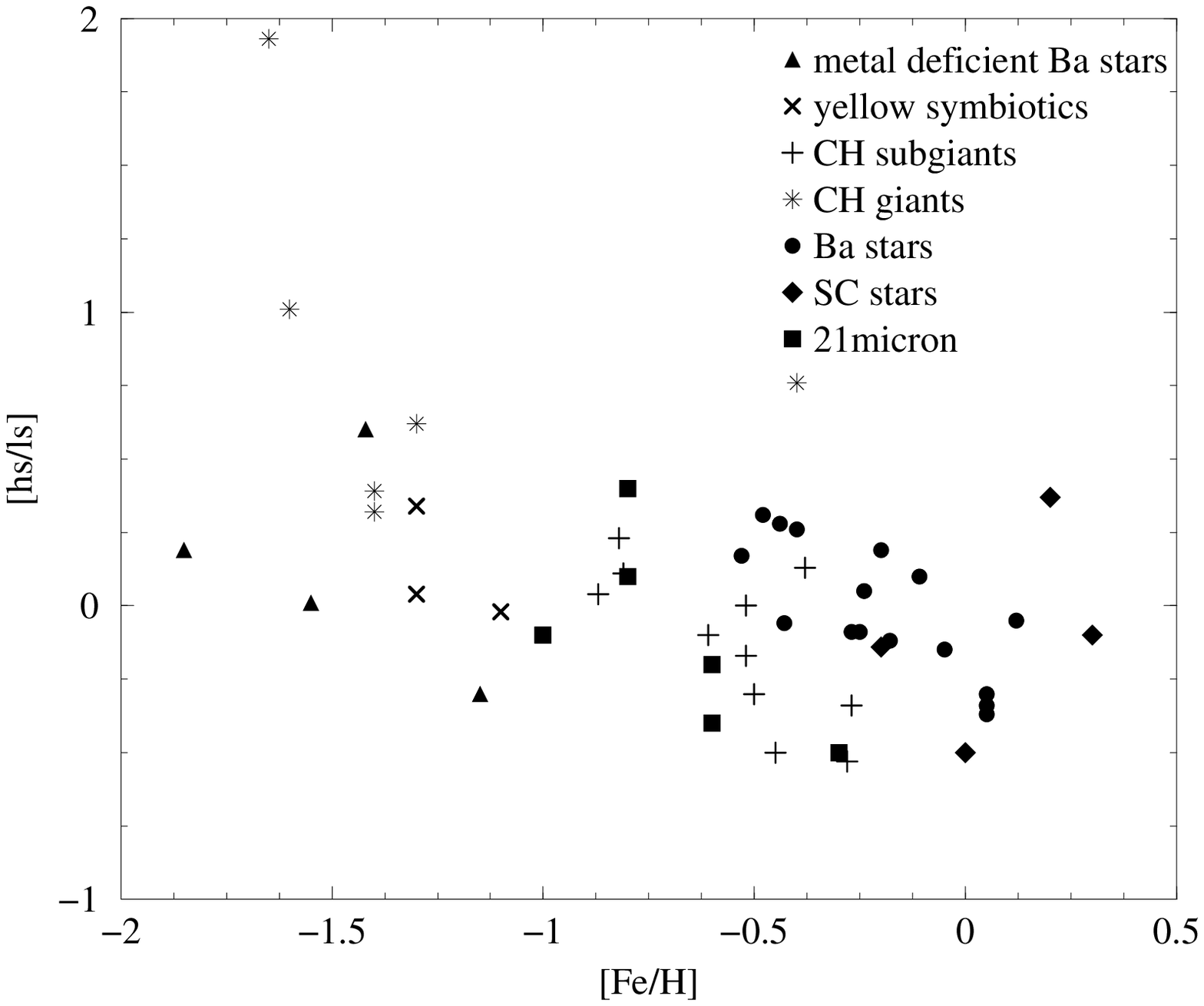}}
\caption{\label{fig:kleurkes}
The [hs/ls] index for different groups of extrinsic (+intrinsic SC) stars, 
compared with the [hs/ls] of the 21\mic\ stars. Data points: see references.}
\end{minipage}
\end{figure*}

To compare the results on the 21\mic\ objects with other stars, we
recalculated all published [hs/ls] indices so that all indices
are calculated using the same species (Y, Zr, Ba, La, Nd, Sm).
The s-process enhancement of the 21\mic\ stars is compared with other
intrinsic objects in Fig.~\ref{fig:handmalaney}. This enhancement
turns out to be larger than in MS, intrinsic S and SC AGB stars and 
comparable with the enhancement in C stars. This result confirms the 
interpretation of the 21\mic\ stars as post-carbon stars and strengthens
the correlation between the C/O ratio (M-MS-S-SC-C sequence) and s-process
enhancement.

A comparison with extrinsic (+intrinsic SC) objects (Fig.~\ref{fig:kleurkes})
shows that the neutron 
exposure of the 21\mic\ stars fall roughly in the same range as the CH
subgiants and is on average lower than in Ba-giants with the same metallicity.
The theoretically expected trend of increasing [hs/ls] index with decreasing
metallicity shows a very large intrinsic spread. Note that the weakly enriched
metal deficient Ba-stars show very small [hs/ls] indices.

\section{Conclusions}
A complete abundance analysis has been carried out on the spectra of six
21\mic\ sources. We confirmed their post-carbon character as they show
high C/O ratios and huge carbon and s-process enhancements. We further 
focussed on the s-process abundance distribution and found a very strong
correlation between the [hs/ls] index (which is a measure for the neutron
nucleosynthesis efficiency) and the total enrichment of the 
s-process elements (which is mainly determined by the dredge-up efficiency).
The anti-correlation of the [hs/ls] index with the metallicity of the 
21\mic\, objects is less well determined and contains a large scatter.
This scatter is certainly intrinsic and confirms that also other fundamental 
parameters strongly determine the internal nucleosynthesis and dredge-up
phenomena during the AGB evolution.

%
%
 \section*{Acknowledgements}
The authors would like to thank the staff of the NTT and WHT telescopes,
the Vienna Atomic Line Database (VALD2), Prof. Kurucz for the distribution of 
his software, Eric Bakker for providing some of the spectra and Christoffel 
Waelkens and Nami Mowlavi for stimulating discussions. 
MR and HVW acknowledge support from the Fund of Scientific Research, Flanders.

%
%
 
\beginrefer
\refer Abia C., Wallerstein G., 1998, MNRAS 293, 89

\refer Busso M., Lambert D.L., Beglio L., et al., 1995, ApJ 446, 775

\refer Decin L., Van Winckel H., Waelkens C., Bakker E.J., 1998, A\&A 332, 928

\refer Luck R.E., Bond H.E., 1991, ApJS 77, 515

\refer Malaney R.A., 1987, Ap\&SS 137, 251 

\refer North P., Berthet S., Lanz T., 1994, A\&A 281, 775

\refer Pereira C.B., Smith V.V., Cunha K., 1998, AJ 116, 1977

\refer Smith V.V., Cunha K., Jorissen A., Boffin H.M.J., 1996, A\&A 315, 179

\refer Smith V.V., Cunha K., Jorissen A., Boffin H.M.J., 1997, A\&A 324, 97

\refer Vanture A.D., 1992, AJ 104, 1997

\endrefer           
\end{document}